\documentclass[12pt]{iopart}

\usepackage{graphicx}
\usepackage{cite}

\begin{document}

\title{Static solutions of Einstein's equations 
        with cylindrical symmetry}

\author{C S Trendafilova and S A Fulling}

\address{Departments of Mathematics and Physics, Texas A\&M University, College Station, TX, 77843-3368}
\eads{\mailto{ryuhime@neo.tamu.edu}, \mailto{fulling@math.tamu.edu}}
\begin{abstract}
In analogy with the standard derivation of the Schwarzschild 
solution, we find all static, cylindrically symmetric solutions 
of the Einstein field equations for vacuum. These include not 
only the well known cone solution, which is locally flat, but 
others in which the metric coefficients are powers of the radial 
coordinate and the space-time is curved.  These solutions appear 
in the literature, but in different forms, corresponding to 
different definitions of the radial coordinate.  Because all the 
vacuum solutions are singular on the axis, we attempt to match 
them to ``interior'' solutions with nonvanishing energy density 
and pressure.  In addition to the well known ``cosmic string'' 
solution joining on to the cone, we find some numerical solutions 
that join on to the other exterior solutions.
\end{abstract}


\section{Introduction}

Static solutions of Einstein's equations with spherical symmetry 
(the exterior and interior Schwarzschild solutions) are staples 
of courses in general relativity. Solutions with cylindrical 
symmetry (combining translation along and rotation around an 
axis) are much less familiar.  In this paper we
construct and study all the vacuum solutions of this sort, and we 
set up the differential equations for nonvacuum solutions and 
find some solutions numerically.
The article is intended for students at the advanced 
undergraduate or beginning 
graduate level; instructors of introductory general relativity 
courses may also wish to use portions of it as a basis 
for exercises appropriate for their classes. The paper is based 
on the undergraduate research thesis of Cynthia Trendafilova 
\cite{thesis}.

In Section 2 we write the most general expression for a 
space-time metric with static and cylindrical symmetry and solve 
the  Einstein vacuum field equations for the components of the 
metric tensor. 
Equivalent solutions may look different because of different 
definitions of the radial coordinate in a cylindrical system.  
At least three different conventions (``radial gauges'') have 
been frequently used in 
the previous literature (see Section~6), none of which is the one 
that seemed most natural to us.
Namely, we adopt the convention that the angular term in the 
metric is $r^2\,\rmd\phi^2$. Then we carefully remove all 
redundant solutions corresponding to the freedom to rescale the 
coordinates.  The principal result is the general solution 
\eref{eq13_1}, in which all the metric components are powers of 
$r$ with exponents related by the constraint \eref{eq11}.

In Sections 3 and 4 we explore the geometrical natures of the 
various solutions obtained. They fall into two classes, the more 
plausible of which shows the circumference $2\pi r$ increasing 
with radial arc length (another natural radial coordinate). 
  Also, 
the transformations into the other radial gauges are worked out; 
in all cases, the metric functions are powers of whatever radial 
coordinate is adopted.  
  Our investigation of the relations among these various 
coordinate conventions  is more systematic than any previous one 
known to us.

The space-times \eref{eq13_1}, which are curved in general, form 
a two-parameter family;
it includes a one-parameter subfamily of 
spaces that are \emph{locally flat}.  When maximally 
continued down to a central singularity, such a space-time is a 
\emph{cone} formed from Minkowski space by removing a wedge, 
characterized by a \emph{deficit angle}.  Although the other 
solutions were discovered in  the early days of general 
relativity \cite{weyl,levi}, they are much less well known than 
the cones.

Like the exterior Schwarzschild solution (when it is not treated 
as a black hole), one expects vacuum cylindrical solutions to be 
physically relevant only over some subinterval of the $r$ axis.  
In Section 5 we search for nonsingular ``interior'' solutions, 
with nonvacuum sources, that can be joined on to the vacuum 
solutions at some radius.  The easiest to find have a locally 
flat cone solution on the outside and a constant energy density 
$\rho$ inside, with pressure $p_z=-\rho$ along the axis and 
vanishing pressures $p_r=p_\phi=0$ in the perpendicular plane;
these are the well known  ``cosmic string'' solutions 
\cite{gott,hisc}.  
Although natural from the point of view of gauge theory 
(see, e.g., \cite{garf}), such 
an equation of state would be surprising for normal matter.  We 
write down the generic Einstein and conservation equations, 
isolate the independent dynamical variables and constraint, and 
construct some numerical solutions with isotropic pressures.  
These solutions   match onto the exterior nonflat vacuum 
solutions.  The results are examples of space-times that also 
deserve to be called ``cosmic 
strings'' but have a more general equation of state for the 
matter inside the string.
We do not claim that these solutions necessarily are 
astrophysically plausible.  
In particular, we have not investigated their 
stability under perturbations that violate the assumed 
symmetries.

In Section 6 we present some historical background on the 
solutions
and gauge choices presented in this paper, while Section 7 offers 
some
concluding remarks.

\section{Vacuum solution of Einstein equations}

In order to solve the Einstein equations for a cylindrically symmetric spacetime, we must first determine the line element with which we would like to work. Let $t$ denote a time coordinate. For a fixed time $t$, a cylindrically symmetric spacetime can be described as follows. There is a central axis of symmetry, with $z$ denoting the coordinate along this axis. Then $\phi$ denotes the coordinate which measures an angle around this axis, and $r$ is the radial coordinate which increases as one moves away from the axis. The line element $\rmd s^2$ is then given by $\rmd s^2=g_{\alpha\beta}\rmd x^{\alpha} \rmd x^{\beta}$, where $g_{\alpha\beta}$ is the metric tensor describing the spacetime. Since we want static, cylindrical symmetry, we require that $\sqrt{-\rmd s^2}$, the clock time measured by a test particle in moving from $(t, r, \phi, z)$ to $(t, r, \phi, z)+(\rmd t, \rmd r, \rmd \phi, \rmd z)$, is independent of $t$, $\phi$, and $z$. It must also be invariant under reversals in the directions or signs of $\rmd t$, $\rmd \phi$, and $\rmd z$. Hence the general static, cylindrically symmetric metric is of the form
\begin{equation}\label{generalmetric}
\rmd s^2=-\rme^{2\Phi}\rmd t^2+\rme^{2\Lambda}\rmd r^2+\rme^{2\Omega}\rmd\phi^2+\rme^{2\Psi}\rmd z^2,
\end{equation}
where $\Phi$, $\Lambda$, $\Omega$, and $\Psi$ are functions of $r$ only. By writing our unknown functions in the form of exponentials, we guarantee that our coefficients will be positive as we would like them to be, and also mirror the standard textbook treatment of the spherically symmetric metric.

In analogy to the standard treatment of spherical symmetry \cite[p 256]{schutz}, we define $r$ so that the coefficient of $\rmd\phi^2$ is equal to $r^2$. We shall call this convention ``tangential gauge." In Sections 3 and 4 we discuss three alternative conventions (out of infinitely many possible), and also the question of whether any generality is lost by this convention. Thus by setting $\rme^{2\Omega}=r^2$ the metric can be written
\begin{equation}\label{eq0}
\rmd s^2=-\rme^{2\Phi}\rmd t^2+\rme^{2\Lambda}\rmd r^2+r^2\rmd\phi^2+\rme^{2\Psi}\rmd z^2,
\end{equation}
where $\Phi$, $\Lambda$, and $\Psi$ are the unknown functions of $r$ for which we would like to solve. The form in which we have written the metric does not restrict the range of $\phi$ to be from 0 to 2$\pi$; instead it runs from 0 to some angle $\phi_\ast$. As we shall show later, $\phi$ can be forced to fill an angle of 2$\pi$ by rescaling $\phi$ and bringing in an additional numerical factor multiplying the angular term, or by also rescaling $r$ and bringing in a numerical factor multiplying the $\rmd r^2$ term.

Using the standard known expressions for the Christoffel symbols, Riemann curvature tensor, and Ricci tensor associated with a given metric \cite[p 134, 159, 164]{schutz}, all of the components of these objects can be calculated for this static, cylindrically symmetric metric. The results for this are presented below.\\
Nonzero Christoffel Symbols:
\begin{eqnarray}
\Gamma^t_{tr}=\Gamma^t_{rt}=\Phi' \nonumber\\
\Gamma^r_{tt}=\Phi'\rme^{2(\Phi-\Lambda)} \nonumber\\
\Gamma^r_{rr}=\Lambda' \nonumber\\
\Gamma^r_{\phi\phi}=-r\rme^{-2\Lambda} \\
\Gamma^r_{zz}=-\Psi'\rme^{2(\Psi-\Lambda)} \nonumber\\
\Gamma^\phi_{r\phi}=\Gamma^\phi_{\phi r} = \frac{1}{r} \nonumber\\
\Gamma^z_{rz}=\Gamma^z_{zr}=\Psi' \nonumber
\end{eqnarray}
Nonzero Riemann Curvature Tensor Components:
\begin{eqnarray}
R^t\!_{\phi\phi t}=r\Phi'\rme^{-2\Lambda} \nonumber\\
R^r\!_{\phi\phi r}=-r\Lambda'\rme^{-2\Lambda} \nonumber\\
R^r\!_{zzr}=(\Psi''+\Psi'^2-\Psi'\Lambda')\rme^{2(\Psi-\Lambda)} \nonumber\\
R^r\!_{ttr}=-(\Phi''+\Phi'^2-\Phi'\Lambda')\rme^{2(\Phi-\Lambda)} \\
R^z\!_{ttz}=-\Psi'\Phi'\rme^{2(\Phi-\Lambda)} \nonumber\\
R^z\!_{\phi\phi z}=r\Psi'\rme^{-2\Lambda} \nonumber
\end{eqnarray}
Nonzero Ricci Tensor Components:
\begin{eqnarray}
R_{tt}=(\Phi''+\Phi'^2-\Phi'\Lambda'+\frac{1}{r}\Phi'+\Psi'\Phi')\rme^{2(\Phi-\Lambda)} \nonumber\\
R_{rr}=-\Phi''-\Phi'^2+\Phi'\Lambda'+\frac{1}{r}\Lambda'-\Psi''-\Psi'^2+\Lambda'\Psi' \nonumber\\
R_{\phi\phi}=r(\Lambda'-\Phi'-\Psi')\rme^{-2\Lambda} \\
R_{zz}=-(\Psi''+\Psi'^2-\Psi'\Lambda'+\Psi'\Phi'+\frac{1}{r}\Psi')\rme^{2(\Psi-\Lambda)} \nonumber
\end{eqnarray}
Primes correspond to differentiation with respect to $r$, e.g., $\Phi'=\frac{\rmd\Phi}{\rmd r}$.

We would like to solve the Einstein field equations for the vacuum solution, which corresponds to $G_{\alpha\beta}=0$. It is easy to show, however, that it is sufficient to calculate the solutions for $R_{\alpha\beta}=0$. We begin with the standard definition of the Einstein tensor,
$G_{\alpha\beta} = R_{\alpha\beta} - \frac{1}{2}Rg_{\alpha\beta}$.
From this we can calculate the trace of the Einstein tensor
$G:=G^\mu\!_\mu = R^\mu\!_\mu - \frac{1}{2}Rg^\mu\!_\mu = R-2R = -R$
and thus obtain the following relation between the Ricci and Einstein tensors:
$R_{\alpha\beta} = G_{\alpha\beta} - \frac{1}{2}Gg_{\alpha\beta}$.
Thus we see that if $R_{\alpha\beta}=0$ then $G_{\alpha\beta}=0$, and conversely, if $G_{\alpha\beta}=0$ then $R_{\alpha\beta}=0$. Thus the solutions to $R_{\alpha\beta}=0$ are also the solutions to the vacuum Einstein field equations, $G_{\alpha\beta}=0$.

By equating the nontrivial components of the Ricci tensor with zero, we obtain a set of four ordinary differential equations for $\Phi$, $\Lambda$, and $\Psi$. We further note that the exponential function is never equal to zero, so the differential equations reduce to
\begin{eqnarray}
(\Phi''+\Phi'^2-\Phi'\Lambda'+\frac{1}{r}\Phi'+\Psi'\Phi')=0, \label{eq1}\\
-\Phi''-\Phi'^2+\Phi'\Lambda'+\frac{1}{r}\Lambda'-\Psi''-\Psi'^2+\Lambda'\Psi'=0, \label{eq2}\\
(\Lambda'-\Phi'-\Psi')=0, \label{eq3}\\
-(\Psi''+\Psi'^2-\Psi'\Lambda'+\Psi'\Phi'+\frac{1}{r}\Psi')=0. \label{eq4}
\end{eqnarray}
We see that \eref{eq3} can be solved for $\Lambda'$ in terms of the other two unknown functions, which can then be substituted into \eref{eq1}, \eref{eq2}, and \eref{eq4} to eliminate $\Lambda'$. Thus this system can be reduced to
\begin{eqnarray}
\Lambda'=\Phi'+\Psi', \label{eq5}\\
\Phi''+\frac{1}{r}\Phi'=0, \label{eq6}\\
\Psi''+\frac{1}{r}\Psi'=0, \label{eq7}\\
\Phi'\Psi'+\frac{1}{r}\Phi'+\frac{1}{r}\Psi'=0. \label{eq8}
\end{eqnarray}
Now \eref{eq5}, \eref{eq6}, and \eref{eq7} are linear second-order equations easily solved by separation of variables, yielding $\Phi=\ln(r^{a_1})+\ln(a_2)$, $\Psi=\ln(r^{b_1})+\ln(b_2)$, and $\Lambda=\ln(r^{a_1+b_1})+\ln(c)$. Also, \eref{eq8} provides the additional constraint that $a_1 b_1 + a_1 + b_1 =0$. Thus the static, cylindrically symmetric metric is
\begin{equation}\label{eq9}
\rmd s^2= -a_2^2r^{2a_1}\rmd t^2 + c^2r^{2(a_1+b_1)}\rmd r^2 + r^2 \rmd\phi^2 + b_2^2r^{2b_1}\rmd z^2
\end{equation}
with $0\leq\phi<\phi_\ast$. The multiplicative constants $a_2$ and $b_2$ can easily be absorbed by rescaling $t$ and $z$, resulting in
\begin{equation}\label{eq10}
\rmd s^2= -r^{2a_1}\rmd t^2 + c^2r^{2(a_1+b_1)}\rmd r^2 + r^2 \rmd\phi^2 + r^{2b_1}\rmd z^2;
\end{equation}
after each change of variables in what follows, we shall carry out this procedure again without comment. Here we have shown that the coefficients must be powers of $r$ as in \eref{eq10}, with $a_1 b_1 + a_1 + b_1 =0$. Since we no longer have to worry about the constants $a_2$ and $b_2$, we now drop the subscripts on $a_1$ and $b_1$, and simply write the constraint as
\begin{equation}\label{eq11}
a b + a + b =0.
\end{equation}
The constant $c$ can also be absorbed by rescaling $r$, which affects the $\rmd\phi^2$ term by bringing out another constant in front, resulting in 
\begin{equation}\label{eq12}
\rmd s^2= -r^{2a}\rmd t^2 + r^{2(a+b)}\rmd r^2 + K^2r^2 \rmd\phi^2 + r^{2b}\rmd z^2.
\end{equation}
This leads to two natural conventions for the $\rmd\phi^2$ term. One can now rescale $\phi$ so that the constant $K^2$ is absorbed, thus redefining the range $\phi_\ast$ of $\phi$. One could instead rescale $\phi$ to fix its range to be from 0 to $2\pi$, in which case the constant remains, multiplying either $\rmd\phi^2$ as in \eref{eq12} or $\rmd r^2$ as in \eref{eq10}. In the work that follows, we use the first convention,
\begin{equation}\label{eq13_1}
\rmd s^2= -r^{2a}\rmd t^2 + r^{2(a+b)}\rmd r^2 + r^2 \rmd\phi^2 + r^{2b}\rmd z^2,
\end{equation}
where the arbitrary constant is hidden in the periodicity, $\phi_\ast$.

\section{Special cases}
\begin{figure}[h]
\begin{center}$
\begin{array}{cc}
\includegraphics[width=7cm]{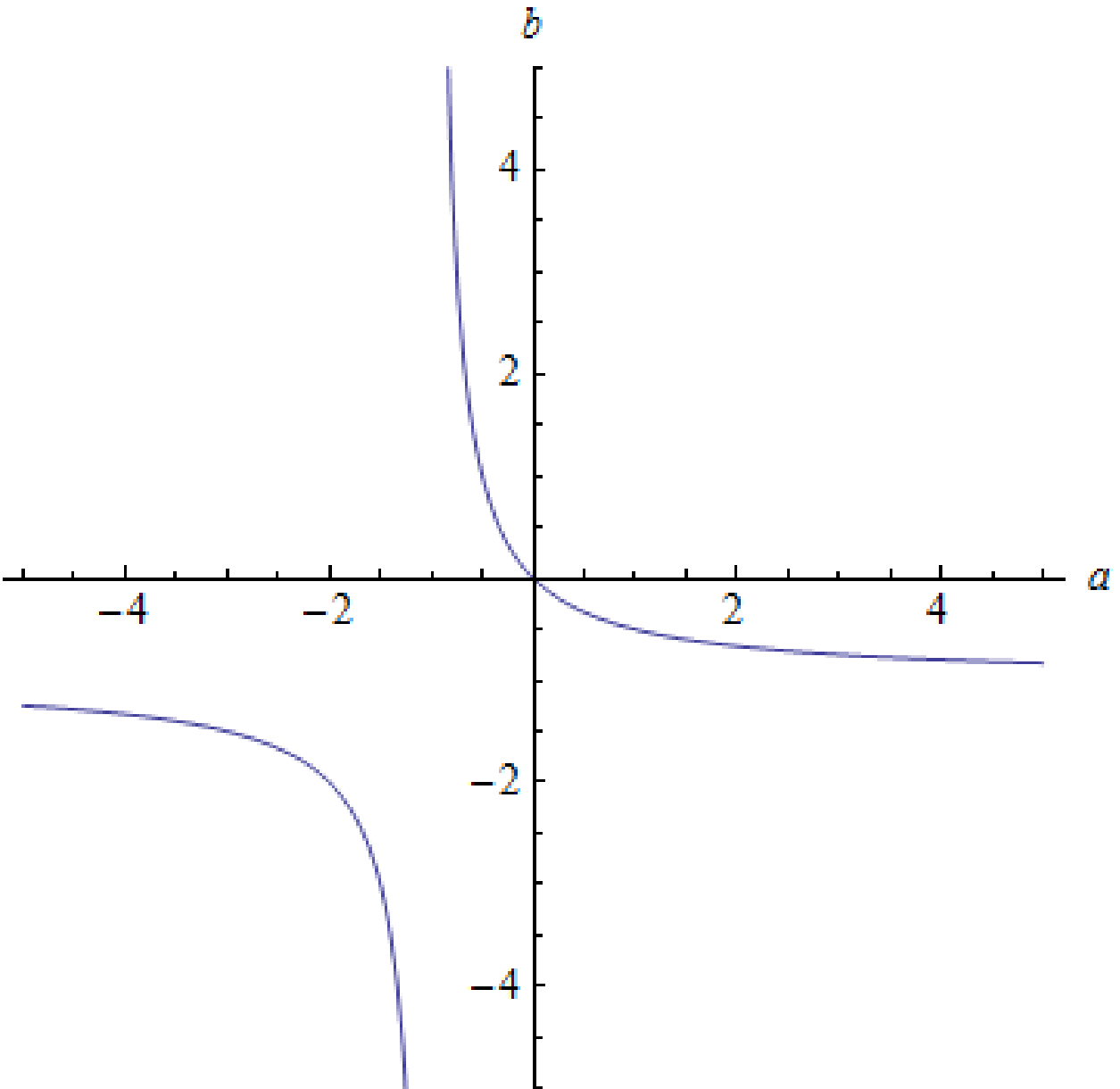} &
\includegraphics[width=7cm]{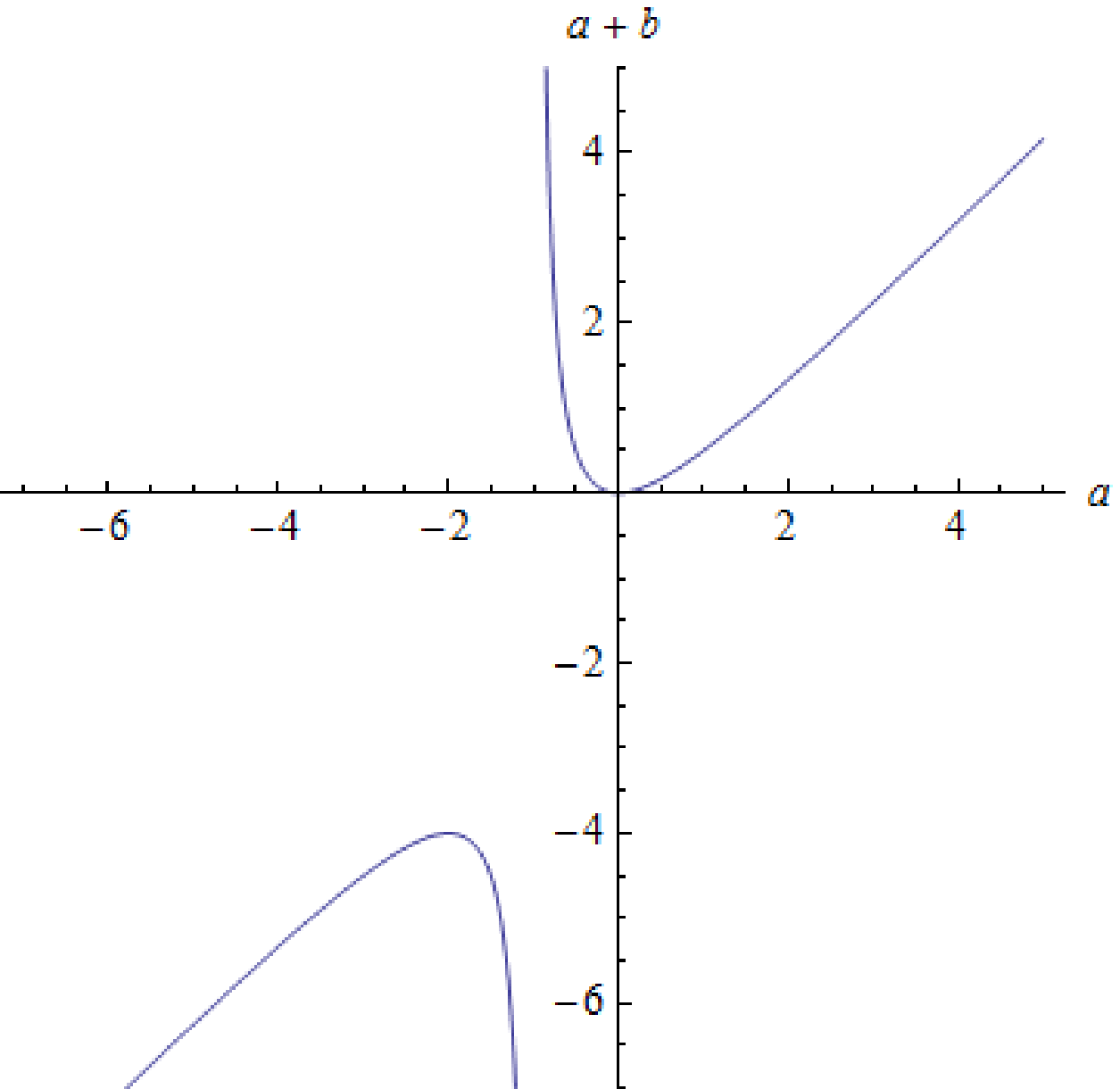}\\
$(a)$ & $(b)$
\end{array}$
\end{center}
\caption{(a) Plot of the relationship between $a$ and $b$. (b) Plot of the relationship between $a$ and $a+b$.}
\label{fig1}
\end{figure}

We now examine in greater detail the relationship between $a$ and $b$, which is illustrated in \fref{fig1}.

We note the existence of several special points on these graphs and examine their significance in various different forms of writing the cylindrical metric.
One such point is $a=b=0$, which reduces the metric of \eref{eq13_1} to
\begin{equation}\label{eq13}
\rmd s^2= -\rmd t^2 + \rmd r^2 + r^2 \rmd\phi^2 + \rmd z^2.
\end{equation}
This describes a cone; that is, flat space missing a wedge of deficit angle $\Delta\phi=2\pi-\phi_\ast$. If $\phi_\ast > 2\pi$, a wedge is added. Ordinary Minkowski space arises as the very special case $\phi_\ast = 2\pi$. It is also useful to note the symmetry between the $t$ and $z$ coordinates, along with the Lorentz symmetry under boosts in the $z$ direction.

Another point of interest is $b=-1$, in which case $a\to\infty$. 
The significance of this 
(apparently singular) case
 can be better demonstrated if we 
rescale $r$, $t$, and $z$ of \eref{eq12} (with the rescaling for 
$r$ given explicitly in Section 4) to write the metric in the 
form
\begin{equation}\label{eq14}
\rmd s^2= -r^{-2b}\rmd t^2 + r^{2(1+b)}\rmd\phi^2 + A^2r^{2b(1+b)}(\rmd r^2+\rmd z^2)
\end{equation}
with $A\equiv c(1+b)$.

If we now treat $A$ as the arbitrary 
constant instead of~$b$, the metric 
remains nonsingular when $b=-1$ in the other terms. After 
rescaling $r$ with 
$\bar{r}=A^{1/[b(1+b)-1)]r}$ to absorb $A$, one gets
\begin{equation}\label{eq15}
\! \rmd s^2= -\bar{r}^{-2b}\rmd\bar{t}^2 + A^{[2(1+b)]/[-b(1+b)-1]}\bar{r}^{2(1+b)}\rmd\phi^2 + \bar{r}^{2b(1+b)}\rmd\bar{r}^2+\bar{r}^{2b(1+b)}\rmd\bar{z}^2.
\end{equation}
We must now rescale $\phi$ as well in order to absorb the final constant in front of the $\rmd\phi^2$ term; this changes the range $\phi_\ast$. The metric becomes
\begin{equation}\label{eq16}
\rmd s^2= -\bar{r}^{-2b}\rmd\bar{t}^2 + \bar{r}^{2(1+b)}\rmd\bar{\phi}^2 + \bar{r}^{2b(1+b)}\rmd\bar{r}^2+\bar{r}^{2b(1+b)}\rmd\bar{z}^2,
\end{equation}
and at the point where $b=-1$ it reduces to
\begin{equation}\label{eq17}
\rmd s^2= -\bar{r}^2\rmd\bar{t}^2 + \rmd\bar{\phi}^2 + \rmd\bar{r}^2+\rmd\bar{z}^2.
\end{equation}
Under the transformation $T=\bar{r}\sinh \bar{t}$ and $R=\bar{r}\cosh \bar{t}$ we see that \eref{eq17} is locally equivalent to flat space,
\begin{equation}\label{eq18}
\rmd s^2= -\rmd T^2 + \rmd\bar{\phi}^2 + \rmd R^2+\rmd\bar{z}^2
\end{equation}
with $\bar{\phi}$ a periodic coordinate.

We also note that the general relationship in \eref{eq11} is symmetric when $a$ and $b$ are switched, corresponding to switching $z$ and $t$. This observation suggests that the case $a=-1$, $b\to\infty$ is parallel to the foregoing one. To see its physical significance, we can write the metric of \eref{eq10} in the form
\begin{equation}\label{eq19} 
\rmd s^2= L^{-2b}r^{2a^2+2a}(-\rmd t^2+\rmd r^2)+L^{2(1+b)}r^{2+2a}\rmd\phi^2 + L^{-2(1+b)}r^{-2a}\rmd z^2
\end{equation}
with $L\equiv[(1+b)/K]^{(1+b)^{-2}}$. At the point where $a=-1$ and $b\to\infty$, this reduces to
\begin{equation}\label{eq20}
\rmd s^2= -\rmd t^2+\rmd r^2+\rmd\phi^2 + r^2\rmd z^2.
\end{equation}
Under the transformation $Z=r\sin z$ and $R=r\cos z$ we get
\begin{equation}\label{eq21}
\rmd s^2= -\rmd t^2 + \rmd\phi^2 + \rmd R^2+\rmd Z^2,
\end{equation}
and this once again looks like flat space locally, but with 
$\phi$ still a periodic coordinate. We note that these locally 
flat solutions are not included in the general solution found in 
Section 2 because there we fixed the coefficient of $\rmd\phi^2$ to 
be $r^2$,
whereas in (\ref{eq18}) and (\ref{eq21}) that coefficient is a 
constant.

\section{Transforming between metric conventions}

The equivalent forms \eref{eq10}, \eref{eq14}, and \eref{eq19} all appear in \cite{mard}. The historical rationale for the last two will be explained in Section 6.

We now examine how transforming from \eref{eq10} to the other 
metric forms affects the radial coordinate $r$. To go from 
\eref{eq10} to \eref{eq14}, we must use $\bar{r}=[c/(a+1)]r^{a+1}$. We see that in this case, the exponent of $r$ is negative whenever 
$a<-1$ or, equivalently, $b<-1$. Under this condition, $r=0$ in 
our gauge choice corresponds to $\bar{r}=\infty$. To go from our form of the metric to that of \eref{eq19}, we require 
$\bar{r}=[c/(b+1)]r^{b+1}$. Once again, the exponent of $r$ is 
negative whenever $b<-1$ ($a<-1$), and in that case $r=0$ 
corresponds to $\bar{r}=\infty$ and vice versa.

Another natural choice for writing the metric, which we shall call ``arc-length gauge," is
\begin{equation}\label{eq22}
\rmd s^2=-A\rmd t^2+B\rmd\phi^2+\rmd r^2+C\rmd z^2,
\end{equation}
where $A$, $B$, and $C$ are once again functions of $r$ only. In 
this 
case, to transform from our gauge to \eref{eq22} we require $\bar{r}=[c/(a+b+1)]r^{a+b+1}$. Under this transformation, the metric becomes
\begin{equation}\label{eq23}
\!\!\!\! \rmd s^2 = - (D\bar{r})^{[2a/(a+b+1)]}\rmd t^2 + \rmd\bar{r}^2 + (D\bar{r})^{[2/(a+b+1)]}\rmd\phi^2 + (D\bar{r})^{[2b/(a+b+1)]}\rmd z^2
\end{equation}
with $D\equiv(a+b+1)/c$. The exponent of $r$ in our definition of $\bar{r}$ is negative whenever $a+b<-1$, which occurs whenever $b<-1$ ($a<-1$). Thus in all three alternate metric forms discussed here, $r=0$ in our gauge corresponds to $\bar{r}=\infty$ in the new gauge whenever $a<-1$ and $b<-1$ (hence $a+b\leq -4$ from \fref{fig1} (b)). The other possibilities have $a+b\geq0$ and $r$ and $\bar{r}$ running in the same direction.

It is also interesting to calculate $W\equiv R^{\alpha\beta\mu\nu}R_{\alpha\beta\mu\nu}$ for the vacuum solution, because this is the simplest nonzero curvature invariant (since $R=0$ from $G_{\alpha\beta}=0$). We get the result that
\begin{equation}\label{eq22_1}
R^{\alpha\beta\mu\nu}R_{\alpha\beta\mu\nu} = 4Cr^{-4(a+b+1)}
\end{equation}
where $C=3a^2+3b^2+3a^2b^2+2ab+2a^2b+2ab^2$. We see that the exponent is negative whenever $a+b>-1$, and in that case, $W \to 0$ as $r \to \infty$. In the other case, $W\to \infty$ as $r \to \infty$ but $W \to 0$ as $\bar{r} \to \infty$, and in that case there is some ambiguity as to which of these limits is the ``outside'' and which is the ``inside''.

\section{Solutions of the Einstein equations with sources}
We would like to find some cylindrical space-times that are not 
singular along the central axis. This requires solving the 
Einstein equations in cases where $T$ has nonzero components. In 
order to proceed with this, we first require a few more basic 
quantities and tensors encountered in general relativity. We 
present here the results for the Ricci scalar, $R$, the Einstein 
tensor, $G_{\mu\nu}$, and the stress-energy tensor, $T_{\mu\nu}$, 
for the cylindrically symmetric metric given in \eref{eq0}.\\
Ricci Scalar:
\begin{eqnarray}
\fl R = \rme^{-2\Lambda}(-2\Phi'' -2\Phi'^2 + 2\Phi'\Lambda\ -2\Psi'' -2\Psi'^2 +2\Psi'\Lambda' -2\Psi'\Phi' + \nonumber\\
\frac{2}{r}\Lambda' -\frac{2}{r}\Phi' -\frac{2}{r}\Psi')
\end{eqnarray}
Nonzero Einstein Tensor Components:
\begin{eqnarray}
G_{tt}=\rme^{2(\Phi-\Lambda)}(-\Psi'' -\Psi'^2 +\Psi'\Lambda' +\frac{1}{r}\Lambda' -\frac{1}{r}\Psi') \nonumber\\
G_{rr}=\Psi' \Phi' +\frac{1}{r}\Phi' +\frac{1}{r}\Psi' \nonumber\\
G_{\phi\phi}=r^2\rme^{-2\Lambda}(\Phi''+\Phi'^2-\Phi'\Lambda'+\Psi''+\Psi'^2-\Psi'\Lambda'+\Psi'\Phi') \\
G_{zz}=\rme^{2(\Psi-\Lambda)}(\Phi''+\Phi'^2-\Phi'\Lambda'-\frac{1}{r}\Lambda' +\frac{1}{r}\Phi') \nonumber
\end{eqnarray}
The components of the stress-energy tensor are defined by setting, for example, $T^r\!_r=p_r$ and lowering an index to get $T_{rr}=p_r \rme^{2\Lambda}$. The other pressure components are defined similarly, and $T^t\!_t=-\rho$.\\
Nonzero Stress Tensor Components:
\begin{eqnarray}
T_{tt}=\rho \rme^{2\Phi} \nonumber\\
T_{rr}=p_r \rme^{2\Lambda} \nonumber\\
T_{\phi\phi}=p_\phi r^2 \\
T_{zz}=p_z \rme^{2\Psi} \nonumber
\end{eqnarray}
From the Einstein field equations in natural units,
 $G_{\mu\nu}=8\pi T_{\mu\nu}$, and the conservation laws, $T^{\alpha\beta}\!_{;\beta}=0$, we get the following five differential equations:
\begin{eqnarray}
0=\frac{\partial{p_r}}{\partial{r}} + p_r (\Phi'+\Psi'+\frac{1}{r}) + \rho\Phi'-p_z \Psi' -\frac{1}{r}p_{\phi}, \label{eq24}\\
4\pi(\rho+p_r+p_{\phi}+p_z)\rme^{2\Lambda} = \Phi''+\Phi'^2-\Phi'\Lambda'+\Psi'\Phi'+\frac{1}{r}\Phi', \label{eq25}\\
\fl 4\pi(\rho+p_r-p_{\phi}-p_z)\rme^{2\Lambda} =  \nonumber\\
-\Phi''-\Phi'^2+\Phi'\Lambda'+\frac{1}{r}\Lambda'-\Psi''-\Psi'^2+\Lambda'\Psi', \label{eq26}\\
4\pi(\rho-p_r+p_{\phi}-p_z)\rme^{2\Lambda} = \frac{1}{r}(\Lambda'-\Phi'-\Psi'), \label{eq27}\\
4\pi(\rho-p_r-p_{\phi}+p_z)\rme^{2\Lambda} = -\Psi''-\Psi'^2+\Psi'\Lambda'-\Psi'\Phi'-\frac{1}{r}\Psi'. \label{eq28}
\end{eqnarray}
We can simplify these by summing \eref{eq25} and \eref{eq26} and subtracting \eref{eq28}. This yields
\begin{equation}\label{eq28_1}
4\pi(\rho+3p_r+p_\phi-p_z)\rme^{2\Lambda} = 2\Psi'\Phi'+\frac{1}{r}\Phi'+\frac{1}{r}\Psi'+\frac{1}{r}\Lambda'.
\end{equation}
We now add and subtract equation \eref{eq27} from \eref{eq28_1}, resulting in
\begin{equation}\label{eq28_2}
4\pi(2\rho+2p_r+2p_\phi-2p_z)\rme^{2\Lambda} = 2\Psi'\Phi' + \frac{2}{r}\Lambda'
\end{equation}
\begin{equation}\label{eq28_3}
4\pi(4p_r)\rme^{2\Lambda} = 2 \Psi'\Phi' + \frac{2}{r}(\Phi'+\Psi').
\end{equation}
We now have a system of differential equations where equations \eref{eq24}, \eref{eq25}, \eref{eq27}, and \eref{eq28} can be solved for $p_r$, $p_\phi$, $p_z$, $\Phi$, $\Psi$, and $\Lambda$ (given $\rho$ and an equation of state relating $\rho$ and the various pressures), and \eref{eq28_3}, which contains only lower-order derivatives of the unknown functions, provides an additional constraint. The system of all five equations is second-order in $\Phi$ and $\Psi$ and first-order in $\Lambda$ and $p_r$.

Differentiating equation \eref{eq28_3} with respect to $r$ and using equations \eref{eq24}, \eref{eq25}, \eref{eq28}, and \eref{eq28_2} to substitute for $p_r'$, $\Lambda'$, $\Phi''$, and $\Psi''$ yields an expression which reduces to $0=0$; thus equation \eref{eq28_3} must hold for all $r$ if it holds at any $r$.

\subsection{Solutions with $\rho=-p_z$, $p_r=p_\phi=0$}
Solving these equations for arbitrary $\rho$, $p_r$, $p_\phi$, and $p_z$ is rather difficult, so a simpler case one can look at is when $\rho=-p_z$ and the other pressure components are zero. In this case, the differential equations reduce to
\begin{eqnarray}
0 = \rho(\Phi'+\Psi'), \label{eq29}\\
0 = \Phi''+\Phi'^2-\Phi'\Lambda'+\Psi'\Phi'+\frac{1}{r}\Phi', \label{eq30}\\
4\pi(2\rho)\rme^{2\Lambda} = \frac{1}{r}(\Lambda'-\Phi'-\Psi'), \label{eq31}\\
0 = -\Psi''-\Psi'^2+\Psi'\Lambda'-\Psi'\Phi'-\frac{1}{r}\Psi', \label{eq32}\\
0 = 2\Phi'\Psi'+ \frac{2}{r}(\Phi'+\Psi'). \label{eq33}
\end{eqnarray}

From equation \eref{eq29} we can see that $\Phi'+\Psi'=0$, allowing us to solve equation \eref{eq31} for $\Lambda' = 8\pi\rho r\rme^{2\Lambda}$, which can easily be solved using integration by parts to get $\Lambda$ for a given $\rho$. From \eref{eq33} and the fact that $\Phi'+\Psi'=0$, we also see that $\Phi'\Psi'=0$. Thus we can conclude that $\Phi'=\Psi'=0$, yielding $\Phi=a_1$ and $\Psi=a_2$ (where $a_1$ and $a_2$ are constants). The metric of the solution can be written as
\begin{equation}\label{gottgeneral}
\rmd s^2 = -\rmd t^2 + \rme^{2\Lambda}\rmd r^2 +r^2\rmd\phi^2 +\rmd z^2.
\end{equation}
This solution (with $\rho = 1/(8\pi r_0^2)$ where $r_0$ is a constant) is widely known; it is usually called the ``cosmic string solution" or ``Gott's solution" \cite{gott,hisc}. Using 
this value of $\rho$, our solutions yield
\begin{equation}\label{eq35}
\rmd s^2 = -\rmd t^2 + [r_0^2/(r_0^2-r^2)]\rmd r^2 + r^2\rmd\phi^2 + \rmd z^2,
\end{equation}
which agrees with \cite{gott} after making the substitution 
$r=\sin(\theta)$ and then rescaling coordinates as necessary. 
This metric is Lorentz-invariant under boosts in the ($z$,$t$) 
plane, and thus $\rho$ and $p_z$ are independent of frame. If we 
did not have the condition that $\rho=-p_z$, the solution would 
not be Lorentz invariant in this way and $\rho$ and $p_z$ would 
not be frame-independent. 
Indeed, generically one would expect the density of the matter in 
a string source to be 
affected by Lorentz contraction when one moves out of the rest 
frame.  But a Gott string, like cosmological dark energy 
(where all components of $p$ equal $-\rho$), has no preferred 
rest frame.  

\subsection{Numerical solutions}
\begin{figure}[h]
\begin{center}$
\begin{array}{cc}
\includegraphics[width=7cm]{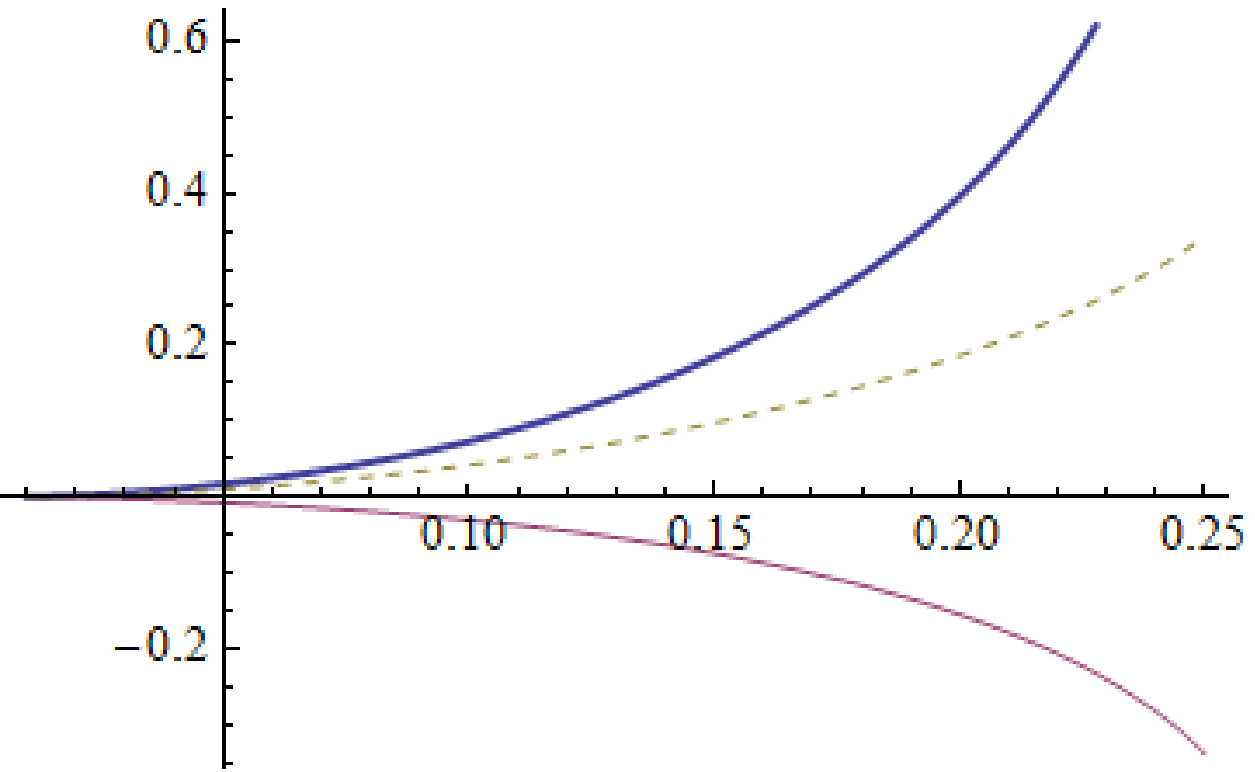} &
\includegraphics[width=7cm]{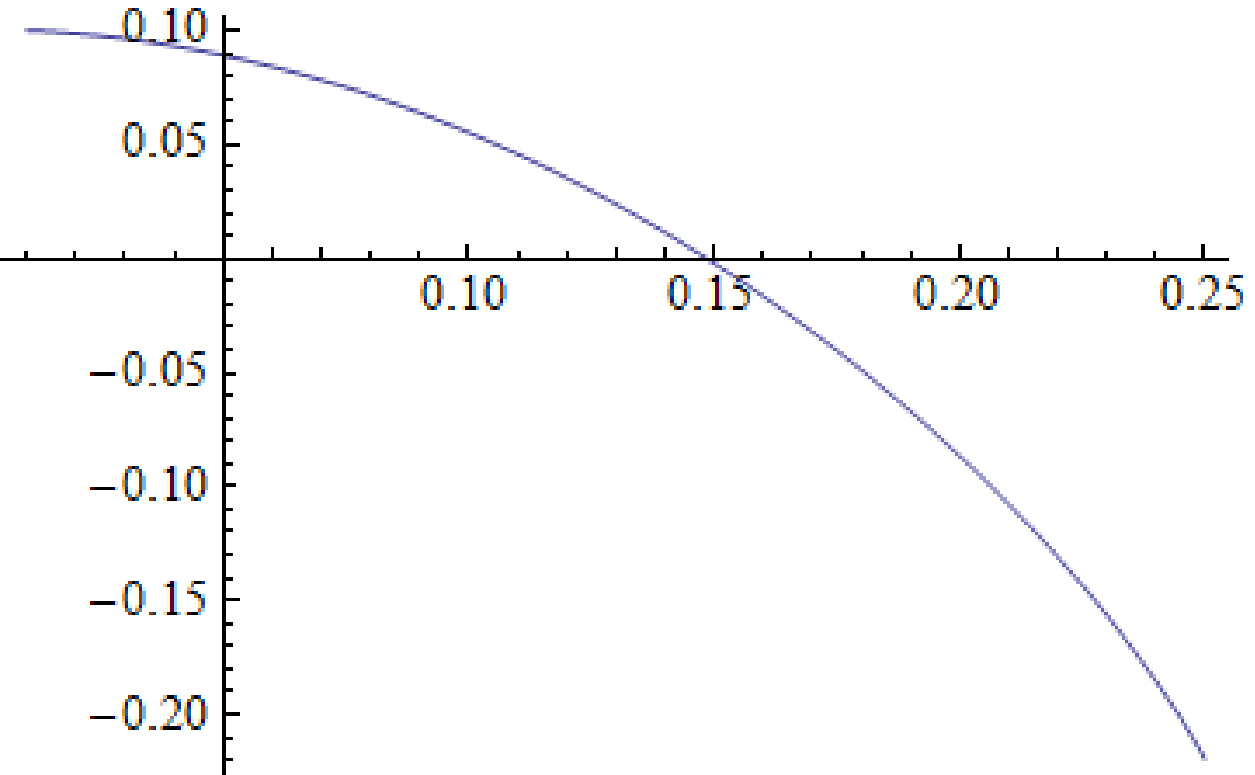}\\
$(a)$ & $(b)$
\end{array}$
\end{center}
\caption{For $\rho=1$ and $p_0=0.1$ (interior): (a) Plot of $\Lambda(r)$ (thick), $\Psi(r)$ (normal), and $\Phi(r)$ (dashed). (b) Plot of $p(r)$.}
\label{fignum1}
\end{figure}

\begin{figure}[h]
\begin{center}$
\begin{array}{cc}
\includegraphics[width=7cm]{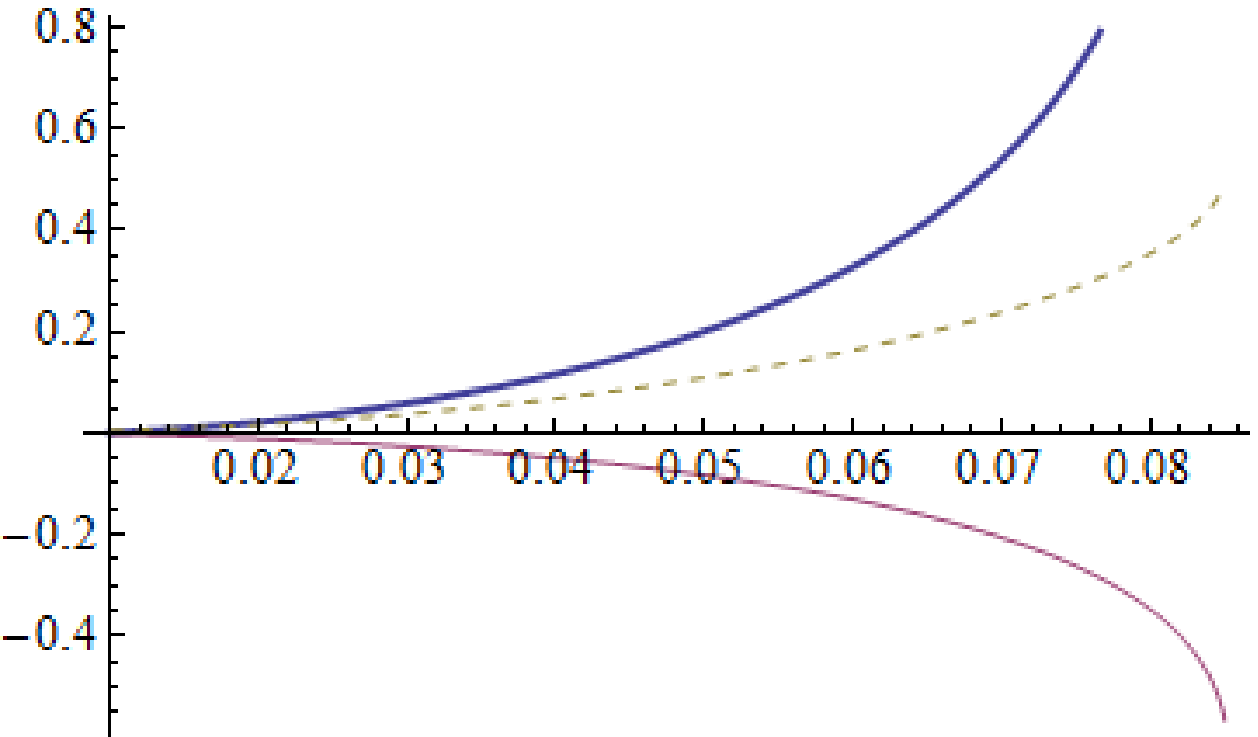} &
\includegraphics[width=7cm]{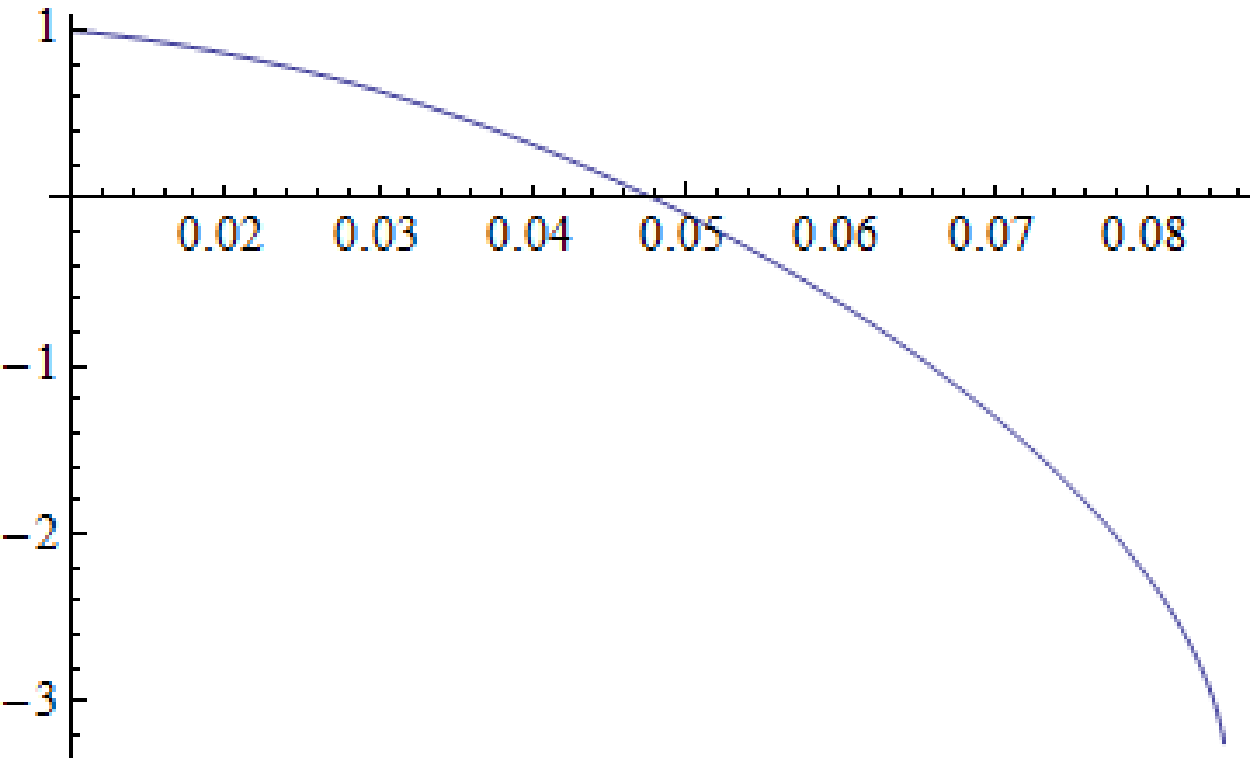}\\
$(a)$ & $(b)$
\end{array}$
\end{center}
\caption{For $\rho=10$ and $p_0=1$ (interior): (a) Plot of $\Lambda(r)$ (thick), $\Psi(r)$ (normal), and $\Phi(r)$ (dashed). (b) Plot of $p(r)$.}
\label{fignum2}
\end{figure}

We now present some numerical solutions (calculated with \textsl{Mathematica}) for the case when $\rho$ is constant out to a radius $R$ and zero outside of this radius, and the pressure is isotropic in all directions, $p_r=p_\phi=p_z \equiv p$. This is analogous to isotropic pressure in the spherically symmetric case. Since our differential equations involve factors of $1/r$, they present problems when trying to solve the system numerically starting from $r=0$. In order to deal with this, we first make power series expansions of $p$, $\Phi$, $\Psi$, and $\Lambda$ around $r=0$. We keep terms up to order $r$ in the $p$ and $\Lambda$ expansions (since our differential equations involve first-order derivatives of these functions) and keep terms up to order $r^2$ in the $\Phi$ and $\Psi$ expansions (since the differential equations involve second-order derivatives of these functions), resulting in
\begin{eqnarray}
p=p_0+p_1 r, \label{t1}\\
\Lambda=\Lambda_0+\Lambda_1 r, \label{t2}\\
\Phi=\Phi_0+\Phi_1 r + \Phi_2 r^2, \label{t3}\\
\Psi=\Psi_0+\Psi_1 r + \Psi_2 r^2 \label{t4}.
\end{eqnarray}
We also take the initial conditions $\Psi=\Phi=\Lambda=0$ at $r=0$ so that the corresponding metric coefficients are equal to $1$ at that point, and choose $\Psi'=\Phi'=0$ to get smooth solutions at the axis.
Equations \eref{t1}--\eref{t4} should satisfy our differential 
equations near $r=0$, so we substitute them into \eref{eq24}, \eref{eq25}, \eref{eq27}, \eref{eq28}, and \eref{eq28_3} (taking $p_r=p_\phi=p_z=p$), and obtain the relationships $p_1=0$, $\Lambda_1=0$, $\Phi_2=\pi(\rho+3p_0)$, and $\Psi_2=-\pi(\rho-p_0)$. After choosing values for $\rho$ and $p_0=p(0)$, we can determine the values of $p$, $\Phi$, $\Psi$, and $\Lambda$ at some small $r$ away from $0$; we take $r=0.01$. We use these as our initial conditions for the numerical calculations and obtain solutions for various values of $\rho$ and $p_0$; two examples of such solutions are provided.
For the case of $\rho=1$, $p_0=0.1$, the results are given in \fref{fignum1}. When $\rho=10$, $p_0=1$, the results are given in \fref{fignum2}. In both cases, the numerical solutions we obtain are not conformally flat (i.e., there are nonzero components of the Weyl tensor).

\subsection{Connecting to exterior solution}
After finding interior solutions numerically for $p_r=p_\phi=p_z=p$, we can then connect them to the exterior vacuum solution found in Section 2. We take $R$ to be the point where $p(r)=0$, and use the values of $\Phi(R)$, $\Phi'(R)$, $\Psi(R)$, $\Psi'(R)$, and $\Lambda(R)$ from the numerical solutions as conditions to determine the unknown coefficients ($a_1$, $a_2$, $b_1$, $b_2$, and $c$) of $\Phi$, $\Psi$, and $\Lambda$ from the vacuum case. We must match our interior solution with the most general vacuum solution from Section 2 which includes all the arbitrary constants, since we chose our initial conditions so that the interior metric coefficients are all $1$ at $r=0$. Because of this, we are not free to scale away $a_2$, $b_2$, and $c$, and we must keep them in the metric in order to match our two sets of solutions.

To perform the matching, we impose the conditions that $\Phi(R)$, $\Phi'(R)$, $\Psi(R)$, $\Psi'(R)$, and $\Lambda(R)$ must be continuous at the boundary where $p=0$. These five continuity conditions provide us with the information needed to find values for the five arbitrary constants from the vacuum solutions. After matching the solutions in this manner, $\Lambda'(R)$ is not necessarily continuous at the boundary. To explain this, it is helpful to look at the metric in the arc-length gauge, given in \eref{eq22}. Obviously one wants $A$, $B$, and $C$ to be continuous so that the metric is continuous. Since $r$ has the same geometrical meaning on both sides of the 
surface, a nonsingular metric should also have continuous $A'$, 
$B'$, and $C'$ (where primes indicate differentiation with 
respect to the radial coordinate in the arc-length gauge).  
Technically, this condition is called ``continuity of the 
second fundamental form'' \cite{herrera2}. Also, in this case, 
the coefficient of $\rmd r^2$ is unity, so it and its derivative are 
automatically continuous. Now letting $\bar{r}$ be the radial coordinate in arc-length gauge, \eref{eq22}, and letting $r$ be the radial coordinate in tangential gauge, \eref{eq0}, we can see that $\frac{\rmd\bar{r}}{\rmd r}=\rme^{2\Lambda}$. Hence, after imposing the conditions that $\Lambda$, $\Phi$, and $\Psi$ must be continuous, continuity of the second fundamental form in arc-length gauge is merely equivalent to continuity of $\frac{\rmd\Phi}{\rmd r}$ and $\frac{\rmd\Psi}{\rmd r}$ in tangential gauge, and the derivative of the coefficent of $\rmd\phi^2$ in tangential gauge is automatically continuous as well. Thus there are no conditions which require that $\frac{\rmd\Lambda}{\rmd r}$ also be continuous at the boundary.

\begin{figure}[h]
\begin{center}$
\begin{array}{c}
\includegraphics[width=7cm]{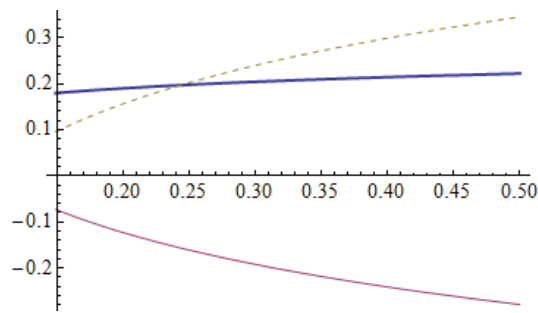} \\
\end{array}$
\end{center}
\caption{For $\rho=1$ and $p_0=0.1$ (exterior): Plot of $\Lambda(r)$ (thick), $\Psi(r)$ (normal), and $\Phi(r)$ (dashed).}
\label{figext1}
\end{figure}

\begin{figure}[h]
\begin{center}$
\begin{array}{c}
\includegraphics[width=7cm]{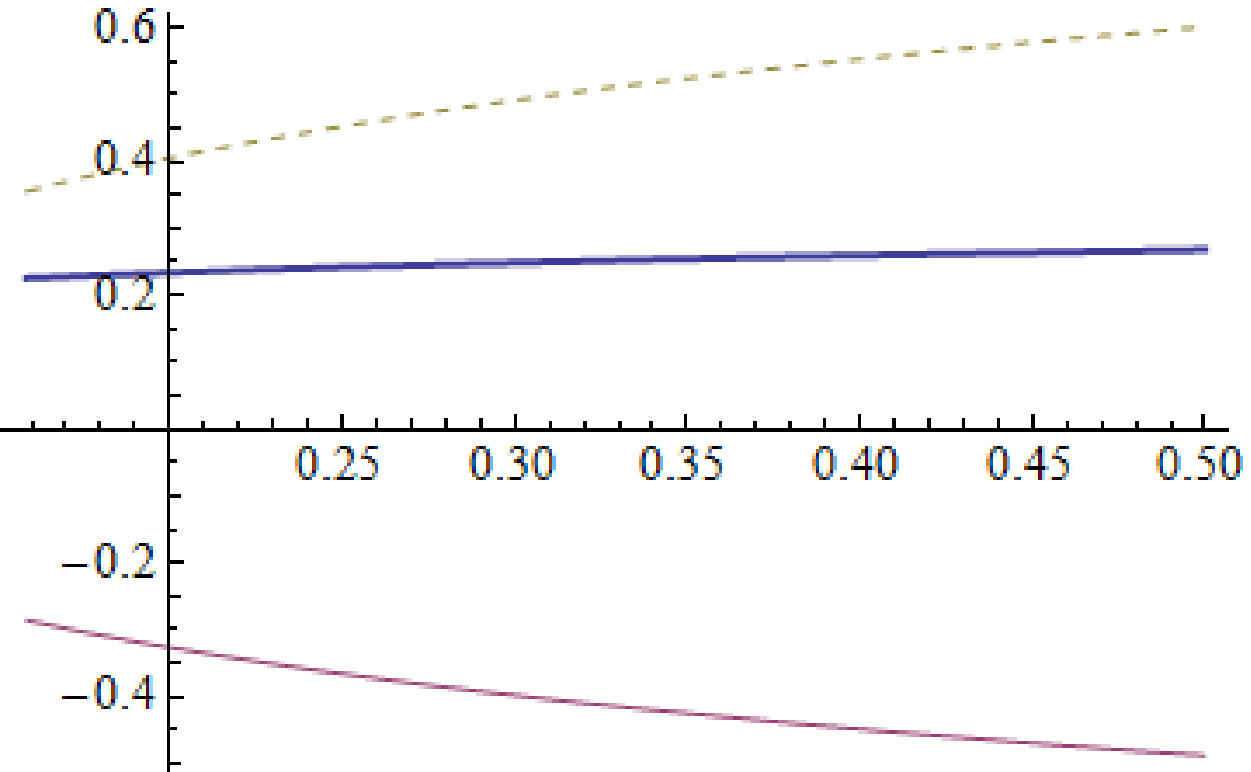} \\
\end{array}$
\end{center}
\caption{For $\rho=10$ and $p_0=1$ (exterior): Plot of $\Lambda(r)$ (thick), $\Psi(r)$ (normal), and $\Phi(r)$ (dashed).}
\label{figext2}
\end{figure}

The results for the two sample cases given above are presented 
in the figures. 
When $\rho=1$, $p_0=0.1$, we get that $R=0.1486$, and the 
resulting exterior solutions are plotted in \fref{figext1}. 
For the case of $\rho=10$, $p_0=1$, we get $R=0.047\,91$, and the 
exterior solutions are plotted in \fref{figext2}. Numerical 
constants for these solutions are given in \tref{exttab}. Also, after rescaling 
coordinates appropriately in order to put the metric in the form 
of equation \eref{eq13_1}, the range of $\phi$ changes as well;
it had to be $2\pi$ for the inner solution to guarantee 
smoothness at 
the origin, and thus the outer solution initially has range 
$2\pi$ when matched with the inner solution. The new value of 
$\phi_*$ is given by $\phi_*=(2\pi)c^{-1/(a_1+b_1+1)}$. The 
values of $\phi_*$ for the two solutions described above are also 
given in the table.

Bi\v{c}\'{a}k et al.\ \cite{bicak} have done a more general study of static perfect-fluid cylinders, including numerical work on incompressible cylinders and their external vacuum solutions. We have compared our numerical results with theirs and found that they agree fairly well. We made comparisons with their parameters $m$ and $\mathcal{C}$, which are related to our notation by $m=-b_1$ and $\mathcal{C} =(2\pi/\phi_*)(a_1+1)^{[-1/(a_1+b_1+1)]}$. The results are presented in \tref{bicaktab} and \tref{bicaktab2}.

In the spherically symmetric case, the Buchdahl theorem \cite[p 269]{schutz}
 requires 
that $R>\frac{9}{4}M$ for any stellar model, where 
$M=\frac{4}{3}\pi\rho R^3$. This implies that 
$R^2\rho<\frac{1}{10}$, where $R^2\rho$ is a dimensionless 
quantity. Although this theorem does not apply to the cylindrically symmetric case, 
it is interesting to note that the inequality does hold in the examples 
studied above.

\begin{center}
\Table{\label{exttab}Numerical constants for exterior solution when $\rho=1$ and $p_0=0.1$ (second column), and $\rho=10$ and $p_0=1$ (third column).}
\br
$R$ & $0.1486$ & $0.047\,91$ \\
$R^2\rho$ & $0.022\,08$ & $0.022\,96$ \\
$a_1$ & $0.2052$ & $0.2136$ \\
$b_1$ & $-0.1703$ & $-0.1761$ \\
$a_2$ & $1.627$ & $2.114$ \\
$b_2$ & $0.6722$ & $0.5435$ \\
$c$ & $1.279$ & $1.342$\\
$\phi_*$ & $4.954$ & $4.732$\\
\br
\endTable
\end{center}

\begin{center}
\Table{\label{bicaktab}Comparison of our numerical results with those of Bi\v{c}\'{a}k et al.\ \cite{bicak}, for $\rho=1$ and $p_0=0.1$.}
\br
 & Bi\v{c}\'{a}k et al. & our results \\
$m$ & $0.1696$ & $0.1703$ \\
$\cal{C}$ & $1.05901$ & $1.05901$ \\
\br
\endTable
\end{center}

\begin{center}
\Table{\label{bicaktab2}Comparison of our numerical results with those of Bi\v{c}\'{a}k et al.\ \cite{bicak}, for $\rho=10$ and $p_0=1$.}
\br
 & Bi\v{c}\'{a}k et al. & our results \\
$m$ & $0.1696$ & $0.1761$ \\
$\cal{C}$ & $1.10063$ & $1.10179$ \\
\br
\endTable
\end{center}

\section{History}
The static, cylindrically symmetric vacuum solutions were found in the early 20th century by Weyl \cite{weyl} and Levi-Civita \cite{levi}.  They were interested in the more general problem of static geometries that are merely \emph{axially} symmetric, that is, depend on $z$ as well as $r$.  It was therefore convenient to use coordinates that treat $r$ and $z$ on the same footing, so the Weyl--Levi-Civita solution was found in the form \eref{eq14}.

Similarly, Marder \cite{mard}, building on work of Rosen \cite{rosen}, studied cylindrically symmetric gravitational waves and hence treated $r$ and $t$ alike, obtaining the static solution in the form \eref{eq19} (four decades after \cite{weyl,levi}).  Marder's paper, which also displays versions \eref{eq10} and \eref{eq14}, was very helpful in the analysis of vacuum solutions in the present paper.

The arc-length gauge \eref{eq22}, leading to solution \eref{eq23}, was probably first used by Evans \cite{evans}; it appears also in much of the cosmic-string literature, such as \cite{garf}, although the Weyl--Levi-Civita convention is also popular there.  Of course, there are many other possible gauge choices; for example, Bi\v{c}\'{a}k et al.\ \cite{bicak} use the convention (in our notation \eref{generalmetric}) $\Phi = -\Lambda$.

During the 1980s, cylindrical solutions received broad attention when the (locally flat) cone solutions were studied as representing the space-time outside a one-dimensional concentration of matter or gauge-field energy, a ``cosmic string''\negthinspace.  Classic papers by Taub \cite{taub} and Vilenkin \cite{vilenkin} considered infinitely thin sources; nonsingular solutions with sources of finite radius were constructed by Gott \cite{gott} and Hiscock \cite{hisc}; and both were studied by many more physicists.  With few exceptions (e.g., \cite{gtrasch,widom}), the possibility that the exterior of a cosmic string might be one of the nonflat Weyl--Levi-Civita solutions was not widely recognized, except when the theory was generalized to include a cosmological constant \cite{linet,tian} (more recently studied in \cite{bicak2}).  Physical enthusiasm for cosmic strings as realistic cosmological objects has diminished in recent years under the pressure of new observational data.

Nevertheless, research on static cylindrical solutions with interior sources and general Weyl--Levi-Civita exteriors has continued and intensified in recent years \cite{herrera2,bicak,evans,bicak2,texeira,bronn,kramer,david1,david2,david3,herrera3,philb,hagg1,hagg2,sharif,herrera1,bicak3,arik,fjal}.  We are not able to provide a complete review of the literature here.

\section{Conclusion}
Cylindrical symmetry in general relativity turns out to be 
similar to spherical symmetry in many ways but quite different in 
others.  The only static vacuum spherical solutions are the 
Schwarzschild metrics parametrized by mass, but, contrary to 
popular belief, there are static vacuum cylindrical solutions 
other than the cone, or cosmic string, spaces parametrized by 
defect angle.  A cone solution is Lorentz-invariant along the 
cylinder axis and hence cannot arise from a matter source 
(at least in the small-radius limit) unless 
the latter has a very unusual equation of state.  We have 
presented numerical solutions with more conventional 
interior sources and more general exterior geometry.

The construction of the various solutions presented here 
illustrates several instructive points.  The choice of gauge 
(coordinate system) is always a major issue in relativity; the 
same space-time can look quite different in different gauges, and 
how (and whether) to choose a standard gauge or ``normal form'' 
for a given problem is not always obvious.  For our problem there 
are several natural ways to fix the gauge, and we have taken 
pains to describe them all and how they are related.  Even after 
a definition of radial coordinate has been selected, further 
steps to a normal form can be taken by linear rescaling of the 
coordinates.  But, as our final calculation shows, sometimes this 
progress must be undone to match solutions for different regions 
properly.

The structure of the Einstein equation system is nontrivial.  
There is one more equation than one might naively expect.  The 
extra equation serves as a constraint on the data.  For the 
cylindrical vacuum solutions this constraint is a simple 
algebraic relation among the parameters, but for our interior 
solutions, extra work was needed to verify that the constraint 
equation is consistent.  The differential orders of the remaining 
equations need to be considered carefully in order to choose the 
correct sort of initial data on the axis and to match the 
interior solution properly to an exterior vacuum solution.

Finally, we observed some surprising ambiguities of 
interpretation.  On the left branch of the vacuum solutions (see 
either half of \fref{fig1}) an increasing radial metric 
component corresponds to decreasing behavior of some of the other 
metric components; therefore, what is naturally considered the 
radial coordinate in our gauge is naturally considered to be the 
reciprocal of such a coordinate in other gauges.  (That is, our 
axis is previous authors' infinity.)  Also, two special 
solutions, \eref{eq13} and \eref{eq21}, are not really 
``cylindrical'' spaces, but rather representations of flat 
space-times in nonstandard coordinate systems.

\ack
We thank David Garfinkle for helpful remarks, and the referee for many valuable pedagogical and bibliographical suggestions, including the list of references at the end of Section 6. Norman Hugh Redington, Davood Momeni, and Luis Herrera also called our attention to additional relevant references in the literature. This work was supported by NSF Grants No. PHY-0554849 and PHY-0968269.

\end{document}